\documentclass[fleqn,10pt]{wlscirep}

\usepackage{multirow}

\title{Tailoring anomalous Nernst effect in stressed magnetostrictive film grown onto flexible substrate}

\author[1]{Ac\'{a}cio Silveira Melo}
\author[1]{Alexandre Barbosa de Oliveira}
\author[1]{Carlos Chesman}
\author[1]{Rafael Domingues Della Pace}
\author[1]{Felipe Bohn}
\author[1,*]{Marcio Assolin Correa}
\affil[1]{Departamento de F\'{i}sica, Universidade Federal do Rio Grande do Norte, 59078-900 Natal, RN, Brazil}
\affil[*]{marciocorrea@dfte.ufrn.br}

\begin{abstract}
The anomalous Nernst effect in nanostructured magnetic materials is a key phenomenon to optimally control and employ the internal energy dissipated in electronic devices, being dependent on for instance the magnetic anisotropy of the active element. 
Thereby, here we report a theoretical and experimental investigation of the magnetic properties and anomalous Nernst effect in a flexible magnetostrictive film with induced uniaxial magnetic anisotropy and under external stress. 
Specifically, we calculate the magnetization behavior and the thermoelectric voltage response from a theoretical approach for a planar geometry and with a magnetic free energy density which takes into account the induced uniaxial and magnetoelastic anisotropy contributions.
Experimentally, we verify modifications of the effective magnetic anisotropy and thermoelectric voltage with the stress and explore the possibility of tailoring the anomalous Nernst effect in a flexible magnetostrictive film by modifying both, the magnetic field and external stress.
We find quantitative agreement between experiment and numerical calculations, thus elucidating the magnetic and thermoelectric voltage behaviors, as well as providing evidence to confirm the validity of the theoretical approach to describe the magnetic properties and anomalous Nernst effect in ferromagnetic magnetostrictive films having uniaxial magnetic anisotropy and submitted to external stress. 
Hence, the results place flexible magnetostrictive systems as a promising candidate for active elements in functionalized touch electronic devices.
\end{abstract}

\begin{document}

\flushbottom
\maketitle

\thispagestyle{empty}

\section*{Introduction}

New and more efficient ways to make use the internal energy converted into heat in electronic devices are of crucial importance for a sustainable future~\cite{SR6p23114, JMMM290p795}. 
In the field of spintronics, the spin caloritronics~\cite{EES7p885910, NMAT11p391399} is an active and exciting area of research with potential to the development of energy-efficient technologies, covering the impacts of a temperature gradient on the production of spin and charge currents in materials~\cite{PRL108p106602}. 

In magnetic materials, the thermoelectric voltage, defined as the direct conversion of the temperature gradient to electrical voltage, is investigated through thermomagnetic phenomena as the spin Seebeck effect (SSE) and the anomalous Nernst effect (ANE)~\cite{APL106p212407, SREP7p6165}. 
Both of them essentially consist in the application of a magnetic field and a temperature gradient, thus generating an electric field $\vec{E}$. 
Specifically for the anomalous Nernst effect in nanostructured magnetic materials, an electric field $\vec{E}_{ANE}$ is induced by the interplay of the magnetization $\vec{m}$ and the temperature gradient $\nabla T$, being a result of $\vec{E}_{ANE} \propto \vec{m} \times \nabla T $.
Thus, given that the corresponding thermoelectric voltage $V_{ANE}$ in magnetic materials is observable even with small thermal gradients at room temperature, ANE becomes a key effect to optimally control and employ the internal energy dissipated in electronic devices.
In the near past, ANE has been extensively investigated, disclosing that $V_{ANE}$ is strongly dependent on the crystallographic orientation in magnetic materials~\cite{APE10p073005}, type of material (for instance, Fe and Ni present ANE coefficient with opposite signs), thickness of the materials in the film geometry~\cite{SREP7p6165, PRB96p174406}, and magnetic anisotropy~\cite{APL106p252405}. 
However, despite the recent advances in this field, there still are several aspects related to the anomalous Nernst effect in nanostructured ferromagnetic materials that are not yet fully understood.

Among the different investigated ferromagnets in the field, materials with high-spin polarization are the main elected to constitute spintronics devices. 
In this context, CoFeB alloys in the film geometry arise as one of the most promising candidates for this kind of application due to its magnetic properties, such as high permeability, high saturation magnetization, low coercive field, and well-defined magnetic anisotropy~\cite{APL105p103504}. 
These features place the CoFeB as alloys suitable to the production of, for instance, magnetic tunnel junctions with high magnetoresistance~\cite{JAP113p213909, JAP100p053903, JAP110p033910, NM3p862} and spin transfer torque~\cite{NM3p862}, as well as to the investigation of a sort of phenomena, as spin pumping~\cite{JAP119p133903, JAP117p163901}, spin Hall effect~\cite{JAP111p07C520}, and inverse spin Seebeck effect~\cite{PRB88p064403}. 
It is known that the magnetic properties of CoFeB films are strongly dependent on the thickness~\cite{NMAT9p721724, JMMM462p2940, PRB83p212404}. 
Films thinner than $5$~nm use to exhibit strong perpendicular magnetic anisotropy (PMA)~\cite{NMAT9p721724}, while the thicker ones commonly show in-plane magnetic anisotropy, which is primarily induced by the formation of pairs of ferromagnetic atoms aligned during the deposition or by oriented binding anisotropy~\cite{JMMM462p2940, PRB83p212404}. 
Moreover, due to its magnetostrictive properties, thick-CoFeB films have been grown on flexible substrates, thus becoming an essential system to the investigation of the influence of stress on the effective magnetic anisotropy~\cite{JMMM426p444, AFM26p4704}, as well as appearing as a promising material for application in flexible magnetic devices~\cite{AIPA6p056106}, disposable electronics, smart cards, light-emitting diodes, wearable electronics and a broad range of sensors~\cite{NATC3p1259, ADVM25p5997, ADVM26p13361342, NAT428p911918, APL105p103504}. 

To pave the way for an effective integration between the thermoelectric voltage experiments and flexible spintronic devices, the anomalous Nernst effect in magnetostrictive films grown onto flexible substrate emerges as a wonderful playground. It occurs not just to the understanding of the fundamental theory of ANE in systems with well-defined magnetic anisotropy, but also to demonstrate the influence of the stress on the effective magnetic anisotropy and thermoelectric effects. 
In this work, we perform a theoretical and experimental investigation of the magnetic properties and anomalous Nernst effect in a flexible magnetostrictive film with induced uniaxial magnetic anisotropy and under external stress. 
Specifically, from the theoretical approach, we calculate the magnetization behavior and the thermoelectric voltage response; experimentally, we verify modifications of the magnetic anisotropy and thermoelectric voltage with the stress.
Thereby, we explore the possibility of tailoring the anomalous Nernst effect in a flexible magnetostrictive film by modifying both, the magnetic field and external stress.
We find quantitative agreement between experiment and numerical calculations, thus elucidating the magnetic and thermoelectric voltage behaviors, as well as providing evidence to confirm the validity of the theoretical approach to describe the magnetic properties and anomalous Nernst effect in ferromagnetic magnetostrictive films having uniaxial magnetic anisotropy and submitted to external stress. 
\section*{Results}

\begin{figure}[!h]\centering
\vspace{.5cm}
\includegraphics[width = 12.5cm]{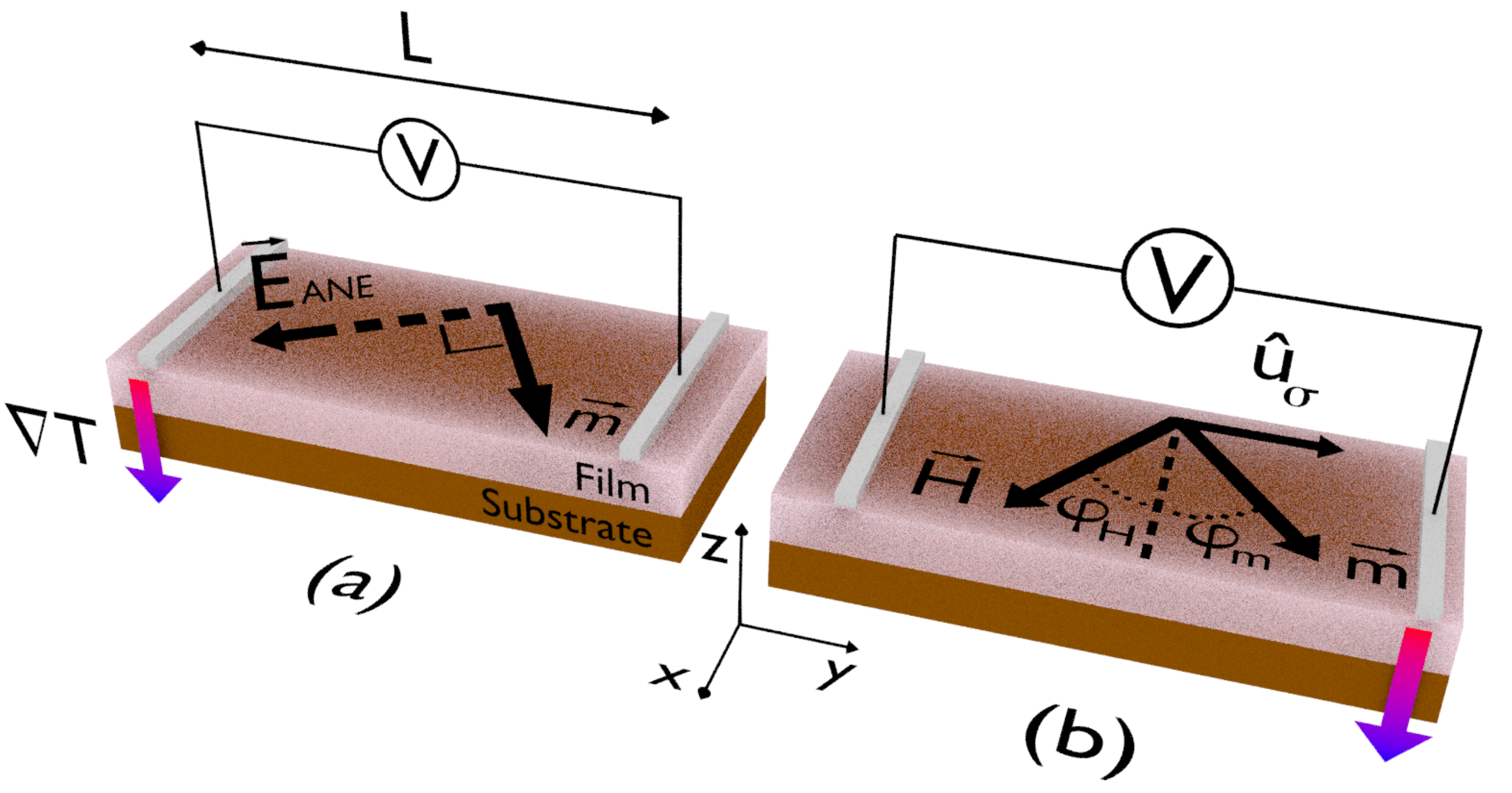} \vspace{-.1cm} 
\caption{{\bf Schematic configuration of our theoretical system --- a ferromagnetic magnetostrictive film having an uniaxial magnetic anisotropy, which is submitted to external stress. } 
{\bf (a)} Anomalous Nernst effect in a film.
In our numerical calculation, while the magnetization $\vec m$ lies in the film plane, the temperature gradient $\nabla T$ is normal to the plane. 
The electric field associated with the anomalous Nernst effect is a result of $\vec{E}_{ANE} \propto \vec{m} \times \nabla T$, and thus the thermoelectric voltage $V_{ANE}$ is proportional to a component of $\vec{E}_{ANE}$, given that its detection is performed with electrical contacts in the ends of the main axis of the film. 
{\bf (b)} Definitions of the vectors and angles employed in the numerical calculations of magnetization and thermoelectric voltage. 
Here we consider the magnetization vector $\vec m$, whose orientation for each given magnetic field value is set by $\theta_m$ and $\varphi_m$, the equilibrium angles with respect to the $z$ and $x$ axes (this latter is also represented by the dashed line above the film), respectively. 
In particular, due to the film geometry and system thickness, here $\theta_m $ is found constantly equal to $90^\circ$. 
The magnetic field is kept in the film plane, i.e., \ $\theta_H = 90^{\circ}$, although its orientation can be modified by varying $\varphi_H$ from $0^\circ$ up to $360^\circ$, having as reference the dashed line indicated in the illustration.
The unit vector $\hat u_k$, not shown here, is defined by $\theta_k$ and $\varphi_k$ and indicates the direction of the uniaxial magnetic anisotropy induced during deposition; 
$\hat u_\sigma$ is given by $\theta_\sigma =90^\circ$ and $\varphi_\sigma =90^\circ$ and describes the direction of the magnetoelastic anisotropy induced by the stress applied to the sample during the experiment; 
and $\hat u_{eff}$, not shown here too, is described by $\theta_{u_{eff}}$ and $\varphi_{u_{eff}}$ and represents the orientation of the effective magnetic anisotropy, a result of the competition between the contributions of the induced uniaxial and magnetoelastic anisotropies. The unit vector normal to the plane of the film, also not shown, is along the $z$ direction, i.e., \ $\hat n = \hat k$.
Finally, for the thermoelectric voltage calculation, the temperature gradient $\nabla T$ is normal to the film plane. 
Further, the $V_{ANE}$ detection is performed with electrical contacts in the ends of the main axis of the film. 
}
\label{Fig_01}
\end{figure}

\noindent{\bf Theoretical approach.} 
Here we focus on a ferromagnetic magnetostrictive film having an uniaxial magnetic anisotropy, which is submitted to external stress; and we model it as a planar system, as illustrated in Fig.~\ref{Fig_01}. 

In the anomalous Nernst effect in magnetic materials, an electric field $\vec E_{ANE}$ is induced by the interplay of the magnetization $\vec m$ of the sample and a temperature gradient $\nabla T$; and the relationship between these quantities can be expressed as 
\begin{equation}
\vec{E}_{ANE}=-\lambda_{N} \mu_{\circ} ( \vec{m} \times {\nabla}T), 
\label{eq_E_ANE}
\end{equation}
\noindent where $\lambda_{N}$ is the anomalous Nernst coefficient and $\mu_{\circ}$ is the vacuum magnetic permeability. 

In our theoretical approach, we consider a typical ANE experiment in a film, in which the temperature gradient is normal to the film plane, while the magnetization lies in the plane, as depicted in Fig.~\ref{Fig_01}(a). 
The corresponding thermoelectric voltage $V_{ANE}$, detected by electrical contacts in the ends of the main axis of the film, is thus given by 
\begin{equation}
V_{ANE} = -\int_{0}^{L}{\vec{E}_{ANE}\cdot d\vec{l}},
\label{eq_V_ANE}
\end{equation}
\noindent where the integration limits are set by the distance between the contacts, which in our case is $L$. 
As a result, the measured $V_{ANE}$ is proportional to the component of $\vec{E}_{ANE}$ along to the direction defined by the contacts, as we can see in Fig.~\ref{Fig_01}(a,b).

For films, $\lambda_{N}$ can be estimated through
\begin{equation}
\label{Nernstcoef}
\lambda_{N} = (V^{Smax}_{ANE} \, t_{f})/(\mu_\circ m_s L \,\Delta T_{f} ), 
\end{equation}
\noindent where $m_s$ is the saturation magnetization of the ferromagnetic alloy, $t_{f}$ is the film thickness, $L$ is the own distance between the probe electrical contacts in the experiment, $\Delta T_{f}$ is the temperature variation across the film (see Methods for details on the $\Delta T_{f}$ estimation and its relation with the experimentally measured temperature variation $\Delta T$ across the sample), and $V^{Smax}_{ANE}$ is a very particular maximum $V_{ANE}$ value that we set from experiment. 
Specifically, this latter is obtained when the sample is magnetically saturated, with $\vec m$ in the film plane and transverse to the detection direction defined by the electrical contacts; this configuration yields a $\vec{E}_{ANE}$ having its higher magnitude and being parallel to the direction of the voltage detection.
After all, our theoretical approach provides a normalized $V_{ANE}$ response, which is rescaled using an experimental $\lambda_N$ value (see Methods for details on the $\lambda_N$ estimation) for comparison. 

It is worth remarking that the amplitude and direction of the magnetization may be changed by the application a magnetic field $\vec H$ and/or stress $\sigma$, thus modifying $\vec E_{ANE}$ and consequently $V_{ANE}$.
To investigate the magnetic properties and thermoelectric voltage response of magnetostrictive films with an induced uniaxial magnetic anisotropy, we employ a modified Stoner-Wohlfarth model~\cite{JMMM453p30}. 
Here, we consider the magnetic free energy density for this system as
\begin{equation} 
\begin{aligned}
\label{eml}
&\xi = \begin{array}{c} - \vec{m} \cdot \vec H +4 \pi m^{2}_{s} \left( \hat{m} \cdot \hat{n} \right) 
- \frac{H_{k}}{2 m_{s}} \left(  \hat{m}\cdot \hat{u}_{k} \right)^2 - \frac{3}{2} \lambda_s \sigma \left( \hat{m} \cdot \hat{u}_{\sigma} \right)^2 \end{array} ,
\end{aligned} 
\end{equation}
\noindent where $\vec m$ is the magnetization vector of the ferromagnetic layer, $\hat m$ is its corresponding versor, $m_s$ is the saturation magnetization, $H_k = 2K_u/m_s$ is the anisotropy field, where $K_u$ is the induced uniaxial magnetic anisotropy constant, $\lambda_s$ is the saturation magnetostriction constant, $\sigma$ is the applied stress, and $\hat n$ is the versor normal to the film plane. 
The theoretical system and the definitions of these important vectors considered to perform the numerical calculations are presented in Fig.~\ref{Fig_01}(b). 
The first term of the magnetic free energy density indicates the Zeeman interaction, the second one is the demagnetizing energy density, and the third term corresponds to the induced uniaxial magnetic anisotropy oriented along $\hat{u}_{k}$. 
Finally, the fourth term describes the magnetoelastic energy density for an elastically medium with isotropic magnetostriction. 
This latter term relates the saturation magnetostriction constant $\lambda_{s}$ and the stress $\sigma$ applied to the system, which give rise to the magnetoelastic anisotropy contribution along $\hat{u}_{\sigma}$. 
In particular, the product between $\lambda_{s}$ and $\sigma$ modifies the effective magnetic anisotropy of the sample. 
In the case of $\lambda_{s}\sigma > 0$, an magnetoelastic anisotropy axis is induced along the same direction of the applied stress, while if $\lambda_{s}\sigma < 0$, this magnetoelastic anisotropy axis is oriented perpendicular to the direction of the stress~\cite{JMMM453p30,Cullity,JMMM420p81}. 

From the appropriate magnetic free energy density, a routine for the energy minimization determines the values the equilibrium angles $\theta_m$ and $\varphi_m$ of the magnetization at a given magnetic field $\vec{H}$, and we calculate the $m/m_{s}$ and the thermoelectric voltage $V_{ANE}$ curves. 

\begin{figure}[!t]\centering
\vspace{.5cm}
\includegraphics[width=10.5cm]{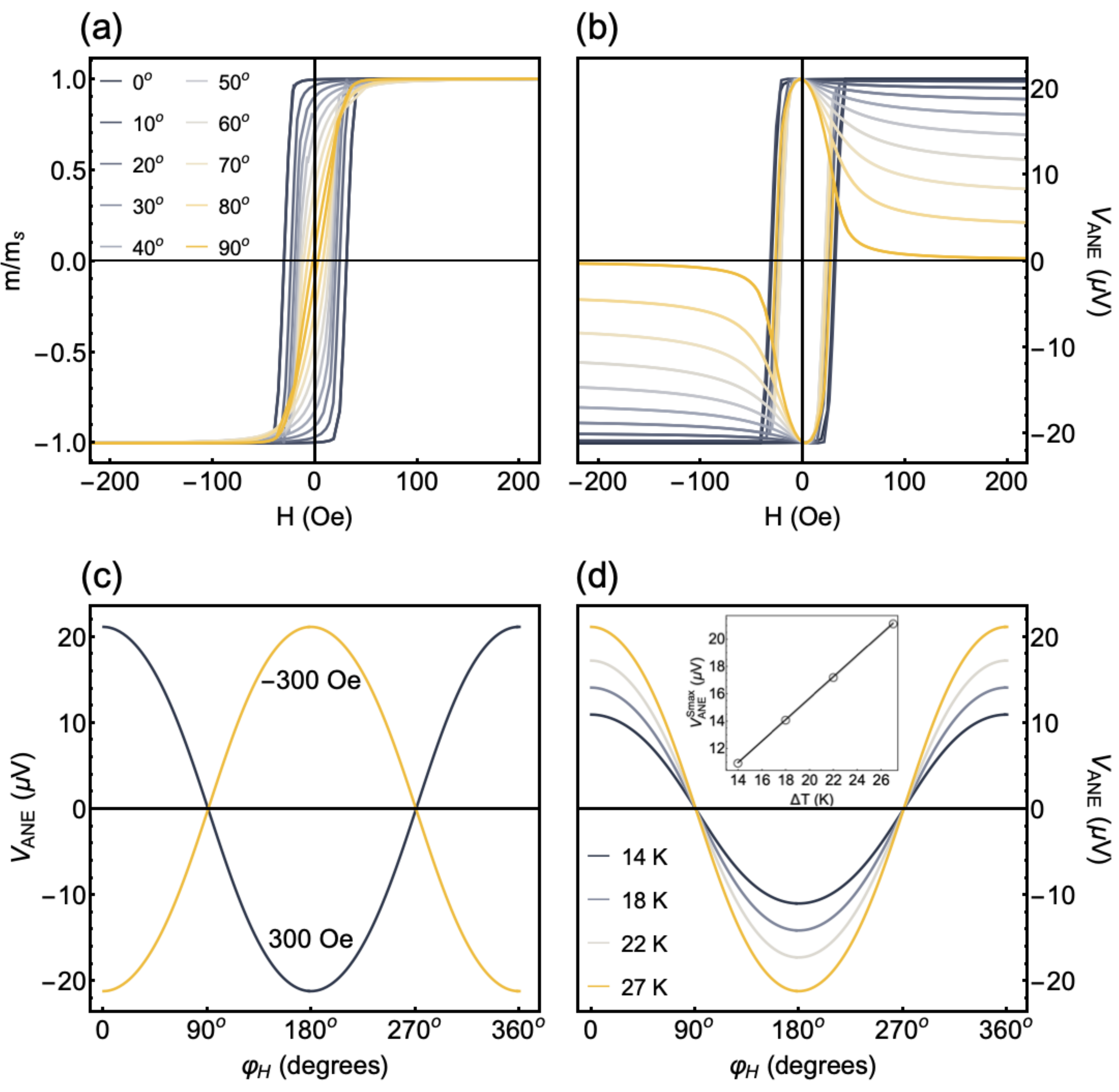} \vspace{-.25cm} 
\caption{{\bf Numerical calculations of the magnetic response and thermoelectric voltage for a film with uniaxial magnetic anisotropy.} 
For the calculations, we consider the following parameters: $m_s= 625$~emu/cm$^3$, $\lambda_{s} = +30.0\times 10^{-6}$, $H_k= 42$~Oe, $\theta_k= 90^{\circ}$, $\varphi_k = 5^{\circ}$, $t_{f}= 300$~nm, $\sigma = 0$~MPa and $\lambda_{N}=16.7\times10^{-6}$~V/KT. 
For the thermoelectric voltage calculation, the temperature gradient $\nabla T$ is normal to the film plane. 
Further, the $V_{ANE}$ detection is performed with electrical contacts in the ends of the main axis of the film. 
Notice that, in this case, this direction is almost transverse to $\hat u_k$, which defines the induced uniaxial magnetic anisotropy.
({\bf a}) Normalized magnetization curves for distinct $\varphi_H$ values. 
({\bf b}) $V_{ANE}$ response, with $\Delta T = 27$~K, as a function of the magnetic field for the very same $\varphi_H$ values. 
({\bf c}) $V_{ANE}$, at $H= \pm 300$ Oe and with $\Delta T = 27$~K, as a function of $\varphi_H$. At this field value, our system is magnetically saturated. 
({\bf d}) Similar plot for the $V_{ANE}$ at $H= + 300$ Oe for distinct $\Delta T$ values. 
In the inset, the dependence of the $V^{Smax}_{ANE}$ value with $\Delta T$. 
}
\label{Fig_02}
\end{figure}

First of all, to confirm the validity of our theoretical approach, we consider a system consisting of a film with uniaxial magnetic anisotropy and without stress, i.e., \ $\sigma = 0$~MPa. 
For the numerical calculations, we take into account the following system parameters: $m_s= 625$~emu/cm$^3$ and $\lambda_{s} = +30.0\times 10^{-6}$, which are characteristic values of the Co$_{40}$Fe$_{40}$B$_{20}$ alloy~\cite{JAP109p07D736,JAP97p10C906,TSF520p2173}, $H_k= 42$~Oe, $\theta_k= 90^{\circ}$, $\varphi_k= 5^{\circ}$ and film thickness of $t_{f}= 300$~nm. 
The magnetic field is in the plane, $\theta_H = 90^{\circ}$, and $\varphi_H$ is varied from $0^\circ$ up to $360^\circ$. 
For the $V_{ANE}$ calculation, we assume $\lambda_{N}=16.7\times10^{-6}$~V/KT and the $\Delta T$ varies between $0$ up to $30$~K. 
To illustrate the results obtained with our routine, Fig.~\ref{Fig_02} presents the magnetization and thermoelectric voltage calculations for this system. 

From Fig.~\ref{Fig_02}(a), magnetization response as a function of the magnetic field for distinct $\varphi_H$ values, the curves clearly present the expected dependence with the orientation between the easy magnetization axis and magnetic field, disclosing all the traditional features of uniaxial magnetic systems~\cite{PTRS240p599}. 
Given that $\varphi_k= 5^{\circ}$, the curve for $\varphi_H = 0^\circ$ reveals the signatures of a typical easy magnetization axis.
Specifically, the magnetization loop presents normalized remanent magnetization close to $1$ and coercive field of around $35$~Oe, this latter close to $H_k$. 
For the magnetization curve at $\varphi_H =90^\circ$ in turn, the normalized remanence reduces to $\sim 0.08$ and the coercive field to around $\sim 2$~Oe, values compatible with a hard magnetization axis. 
Remarkably, the small deviation of $\varphi_k$ yields an hysteretic behavior for $\varphi_H =90^\circ$, given rise to a typical hard axis behavior with a non-zero coercive field.

Regarding the thermoelectric voltage calculation, Fig.~\ref{Fig_02}(b) shows the $V_{ANE}$ response, with $\Delta T = 27$~K, as a function of the magnetic field for distinct $\varphi_H$ values. 
Notice the quite-interesting evolution in the shape of the curves as the magnitude and orientation of the field are altered. 
It is worth remarking that the $V_{ANE}$ detection is performed with electrical contacts in the ends of the main axis of the film, and here this direction is almost transverse to $\hat u_k$, which defines the orientation of the induced uniaxial magnetic anisotropy.
At small $\varphi_H$ values, the $V_{ANE}$ curves seem to mirror the magnetization loops.
This feature is a signature that the magnetization is primarily kept close to the easy magnetization axis, thus favoring the alignment between $\vec{E}_{ANE}$ and the detection direction defined by the electrical contacts. 
Therefore, $V_{ANE}$ is directly proportional to the magnitude of the magnetization. 
On the other hand, as the $\varphi_H$ value increases and the field is not nearby the easy magnetization axis, the $V_{ANE}$ curves lose the squared shape. 
This fact is caused by the competition between two energy terms, the first associated with the Zeeman interaction and the second one related to the induced uniaxial magnetic anisotropy. 
At the low magnetic field range, the magnetization remains close to the easy magnetization axis, leading to high $V_{ANE}$ values. 
However, as the magnetic field increases, the Zeeman energy term dominates, and the magnetization rotates out from the easy axis, following the field.
As a result, the component of $\vec{E}_{ANE}$ along the detection direction defined by the electrical contacts decreases and $V_{ANE}$ is drastically reduced.

Looking at the themoelectric voltage at a specific magnetic field, Fig.~\ref{Fig_02}(c) presents the $V_{ANE}$ values, at $H = \pm 300$~Oe and with $\Delta T = 27$~K, as a function of $\varphi_H$. 
At the field value of $\pm 300$~Oe, our system is magnetically saturated, in a sense that the magnetization follows the orientation of the magnetic field. 
Noticeably, the curves draw a clear dependence of $V_{ANE}$ with the sign of the magnetic field and primarily with $\varphi_H$, this latter evidenced by a well-defined cosine shape, as expected.
From Eqs.~(\ref{eq_E_ANE}) and (\ref{eq_V_ANE}), and taking into account the definitions illustrated in Fig.~\ref{Fig_01}(a), one may notice that $V_{ANE}$ relates with $|\vec m|$ and $\varphi_H$ through
\begin{equation}
V_{ANE} = \lambda_N \mu_0 |\vec m| |\nabla T| L \cos \varphi_m.
\label{eq_V_ANE_component}
\end{equation}
Therefore, given that here $\varphi_H = \varphi_m$, our numerical findings in Fig.~\ref{Fig_02}(c) are consistent with the $V_{ANE}$ angular dependence expressed in Eq.~(\ref{eq_V_ANE_component}).

Further, Fig.~\ref{Fig_02}(d) shows the $V_{ANE}$ behavior as a function of $\varphi_H$ at $H = + 300$~Oe for selected $\Delta T$ values. 
In this case, although the amplitude of the curves is altered, its angular dependence is kept constant, also corroborating Eq.~(\ref{eq_V_ANE_component}). 
The maximum $V_{ANE}$ value at this magnetic saturation state is found for $\varphi_H = 0^\circ$, which corresponds to $V^{Smax}_{ANE}$ according to our setup illustrated in Fig.~\ref{Fig_01}(a,b). 
Moreover, a linear dependence of $V^{Smax}_{ANE}$ with $\Delta T$ is found, as we can see in the inset.

After all, the numerical calculations obtained with our theoretical approach allows us to have an overview of the ANE for a nanostructured system consisting of a film with an induced uniaxial magnetic anisotropy. 
From now, to go beyond, the main challenge to the description of different systems resides in the writing of the appropriate magnetic free energy density. 

\vspace{.3cm}
\noindent{\bf Comparison with the experiment.} 
The previous calculations have qualitatively described the main features of magnetization and thermoelectric voltage in a film with uniaxial magnetic anisotropy and without stress. 
From now on, we investigate the quasi-static magnetic properties and anomalous Nernst effect in Co$_{40}$Fe$_{40}$B$_{20}$ (from now on called CoFeB) films with thickness of $300$~nm grown onto rigid (glass) and flexible (Kapton$^{\textrm{\tiny \textregistered}}$) substrates (see Methods for details on the films and experiments). 
In addition, for the flexible films, we perform experiments for the sample with and without external stress. 

\begin{figure}[!h]\centering
\vspace{.5cm}
\includegraphics[width=10.5cm]{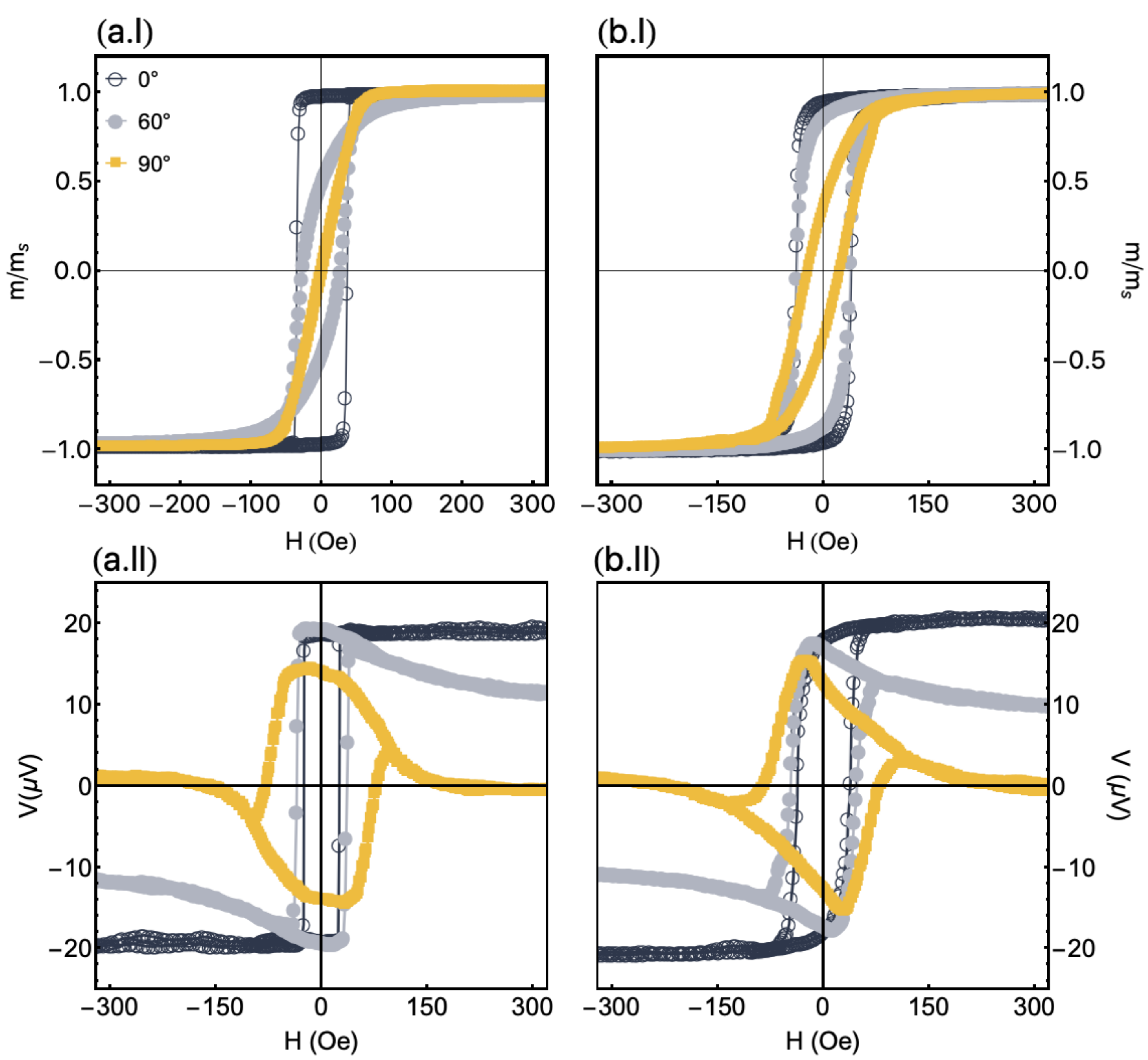} \vspace{-.1cm} 
\caption{{\bf Experimental results of the magnetic response and thermoelectric voltage for CoFeB films without stress.} 
({\bf a}) Normalized magnetization curves and $V_{ANE}$ response at selected $\varphi_H$ values for the CoFeB film grown onto rigid substrate. ({\bf b}) Similar plot for the CoFeB film grown onto flexible substrate. The thermoelectric voltage measurements are performed with $\Delta T = 27$~K.
}
\label{Fig_03}
\end{figure}

Figure~\ref{Fig_03} shows experimental results of the magnetic response and the thermoelectric voltage for the CoFeB films without stress.
From Fig.~\ref{Fig_03}(a.I), magnetization response for the film grown onto rigid substrate as a function of the magnetic field for distinct $\varphi_H$ values, the curves assign the characteristic behavior of a system with well-defined uniaxial magnetic anisotropy, with easy magnetization axis along $\varphi_k \approx 0^\circ$, i.e., \ aligned perpendicularly to the main axis of the sample. 
In particular, these experimental data are in quantitative agreement with the numerical calculations previously shown in Fig.~\ref{Fig_02}(a), disclosing coercive field of $\sim 40$~Oe and normalized remanent magnetization of $\sim 1$ for $\varphi_H=0^\circ$, while values close to zero for both at $\varphi_H=90^\circ$. 

Regarding the thermoelectric voltage results, Fig.~\ref{Fig_03}(a.II) shows the $V_{ANE}$ response for the film grown onto a rigid substrate, with $\Delta T = 27$~K, as a function of the magnetic field for distinct $\varphi_H$ values.
The experimental $V_{ANE}$ curves are also in concordance with the numerical calculations presented in Fig.~\ref{Fig_02}(b), reflecting all features of systems with uniaxial magnetic anisotropy and without stress. 
Specifically, we find here the evolution in the shape of the curves as the magnitude and orientation of the field are altered.
Moreover, at high magnetic field values, when the film is magnetically saturated, we clearly verify the angular dependence of $V_{ANE}$, having $V^{Smax}_{ANE} \approx 20\,\mu$V, and with the characteristic reduction of its value to zero as $\varphi_H$ increases to $90^\circ$.
However, it is worth remarking that, at the low field range, a discrepancy between theory and experiment may be found. 
In particular, as $\varphi_H$ is raised to $90^\circ$, $V_{ANE}$ at low fields do not reach the maximum of $20\,\mu$V found at $\varphi_H = 0^\circ$ and expected for this sample under these experimental conditions, a value also evidenced numerically in Fig.~\ref{Fig_02}(b). 
This difference is primarily associated with magnetization process at low magnetic fields and to the existence of magnetic domains in the film, a fact that is not taken into account in our microspin modified Stoner-Wohlfarth theoretical approach. 
As a result, the magnitude of the magnetization in the experiment is not kept constant at the low-field levels, leading to a reduction of the $V_{ANE}$ value.

We find similar behaviors of the magnetization and thermoelectric voltage for the film grown onto rigid and flexible substrates, as we can see in Fig.~\ref{Fig_03}(a,b).
However, although the flexible film has in-plane uniaxial magnetic anisotropy, the magnetization curves suggest the existence of anisotropy dispersion. 
In this case, while coercive field $\sim 42$~Oe and remanent magnetization of $\sim 1$ are found for $\varphi_H=0^\circ$, respective values close to $30$~Oe and $0.5$ are measured at $\varphi_H=90^\circ$. 
These values found for $\varphi_H=90^\circ$ are indicators of the magnetic anisotropy dispersion, which is primarily related to the internal stress stored in the film during the growth onto the flexible substrate. 
Although the thermoelectric voltage response presents all the expected general features, this dispersion is also the responsible by the small modifications in the shape and amplitude of the $V_{ANE}$ curves, when compared to the results measured for the film grown onto the rigid substrate.

From the experimental results found for rigid and flexible films, shown in Fig.~\ref{Fig_03} and compared with the calculated ones presented in Fig.~\ref{Fig_02}, we corroborate the angular dependence of the magnetization and thermoelectric voltage curves in films with uniaxial magnetic anisotropy.
Going beyond, from now on, we focus our efforts on the magnetic response and thermoelectric voltage in a magnetostrictive film under stress. 
As aforementioned, CoFeB alloy has high saturation magnetostriction constant, reaching $\lambda_{s} \approx +30.0\times 10^{-6}$~\cite{JAP97p10C906,TSF520p2173}. 
Therefore, we are able to manipulate here the magnetic properties, i.e.~the effective magnetic anisotropy, of our magnetostrictive CoFeB film grown onto a flexible substrate through the application of stress.

\begin{figure}[!h]\centering
\vspace{.5cm}
\includegraphics[width=10.5cm]{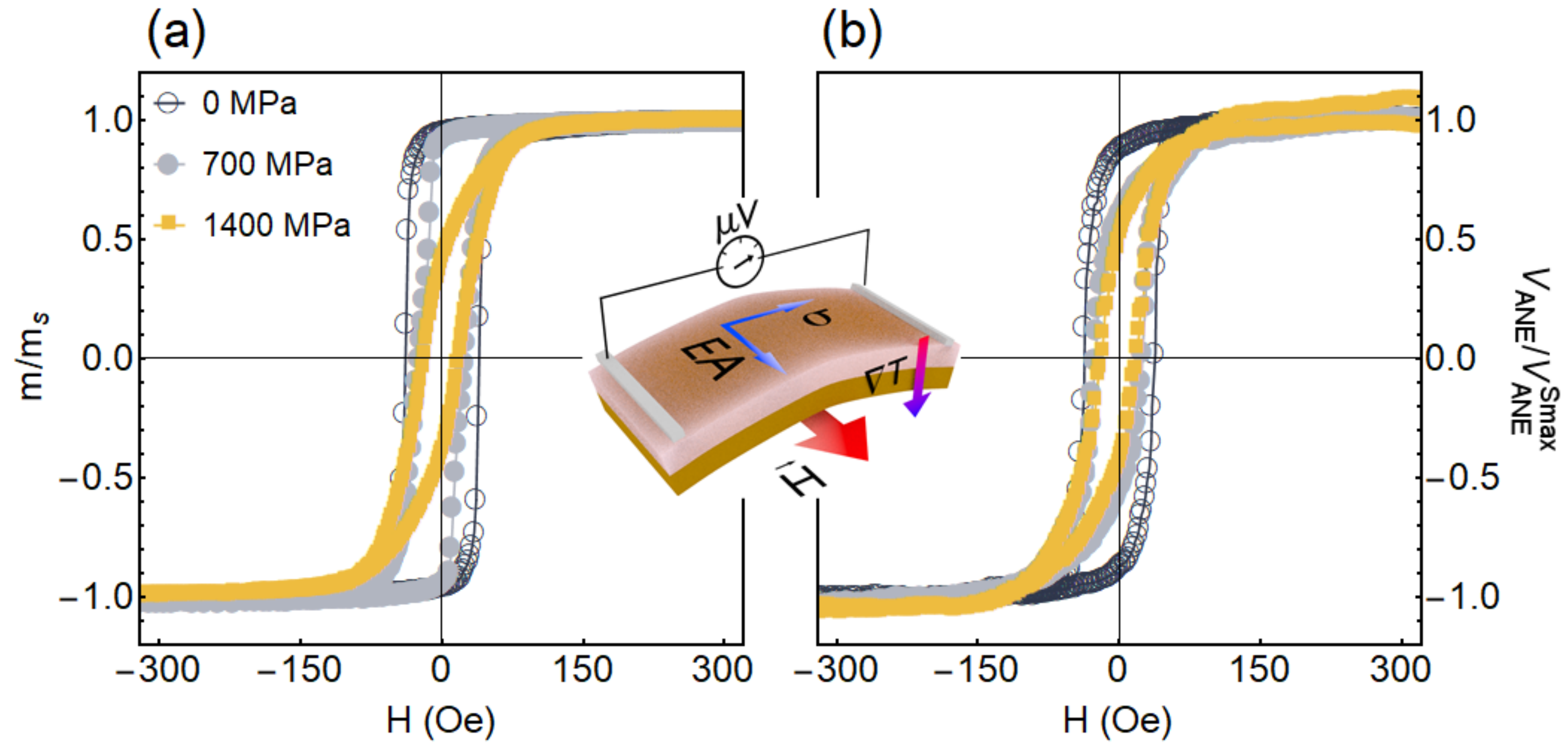} \vspace{-.1cm} 
\caption{{\bf Experimental results of the magnetic response and thermoelectric voltage for the flexible CoFeB film under stress.}
({\bf a}) Normalized magnetization curves and ({\bf b}) normalized $V_{ANE}$ response for the flexible CoFeB film under distinct stress values. 
The measurements are performed with $\varphi_H = 0^\circ$, and bending the sample along its main axis.
Moreover, we set $\Delta T = 27$~K for the $V_{ANE}$ acquisition. 
The schematic configuration of the sample in the ANE experiment uncover the easy magnetization axis (EA) induced during deposition, the orientation of the stress $\sigma$ in the bent film, as well as the directions of the magnetic field, temperature gradient and $V_{ANE}$ detection.}
\label{Fig_04}
\end{figure}

Figure~\ref{Fig_04} shows experimental results of the normalized magnetization curves and normalized $V_{ANE}$ response for the flexible CoFeB film under distinct stress values. 
The measurements are performed at $\varphi_H = 0^\circ$, with the stress along the main axis of the sample (see Methods for details on the stress application). 
Given that the saturation magnetostriction constant is positive, the product $\lambda_{s}\sigma > 0$ for positive stress, thus yielding a magnetoelastic anisotropy axis oriented along the main axis of the sample, i.e.~$\theta_\sigma = 90^\circ$ and $\varphi_\sigma = 90^\circ$. 

From the magnetization response shown in Fig.~\ref{Fig_04}(a), we uncover the modification of the effective magnetic anisotropy with the stress level. 
Specifically, while the effective anisotropy is well-described by $\varphi_{k_{eff}}$ roughly close to $\varphi_k$ for $\sigma = 0$, the changes in the coercive field, remanent magnetization as well as the own shape of the magnetization curve as the stress level increases are a straight consequence of the raise of the magnetoelastic anisotropy contribution to the effective magnetic anisotropy. 
This raise leads to changes in the orientation $\hat u_{k_{eff}}$ of the effective magnetic anisotropy with the increase of the stress value, thus modifying the resulting whole magnetic behavior and magnetization curves. 
For $\sigma = 700$~MPa, the squared curve with smaller coercive field reveals an intermediate magnetic behavior, arisen from an effective magnetic anisotropy that is still roughly close to the direction of the magnetic field at $\varphi_H = 0^\circ$, but with a significant component along the main axis of the sample due to the magnetoelastic contribution. 
However, for $\sigma = 1400$~MPa, the stress is high enough to set the effective magnetic anisotropy axis close to $\hat u_\sigma$, i.e.~$\hat u_{eff} \approx 90^\circ$. 
Consequently, as expected, the magnetization curve at $\varphi_{H} = 0^\circ$ has fingerprints of a hard magnetization axis, with non-zero coercive field and normalized remanent magnetization due to anisotropy dispersion.
Therefore, the whole magnetic behavior of the flexible magnetostrictive film under stress is a result of the competition between the induced uniaxial magnetic anisotropy and the magnetoelastic anisotropy contribution. 
As a consequence, this competition allows us to tailor the effective magnetic anisotropy and, consequently, the anomalous Nernst effect. 
Remarkably, as we can corroborate from Fig.~\ref{Fig_04}(b), modifications in the $V_{ANE}$ response are already visible at $\varphi_H = 0^\circ$ when just the stress is altered in an experimental manner.

\begin{figure}[!h]\centering
\vspace{.5cm}
\includegraphics[width=15.5cm]{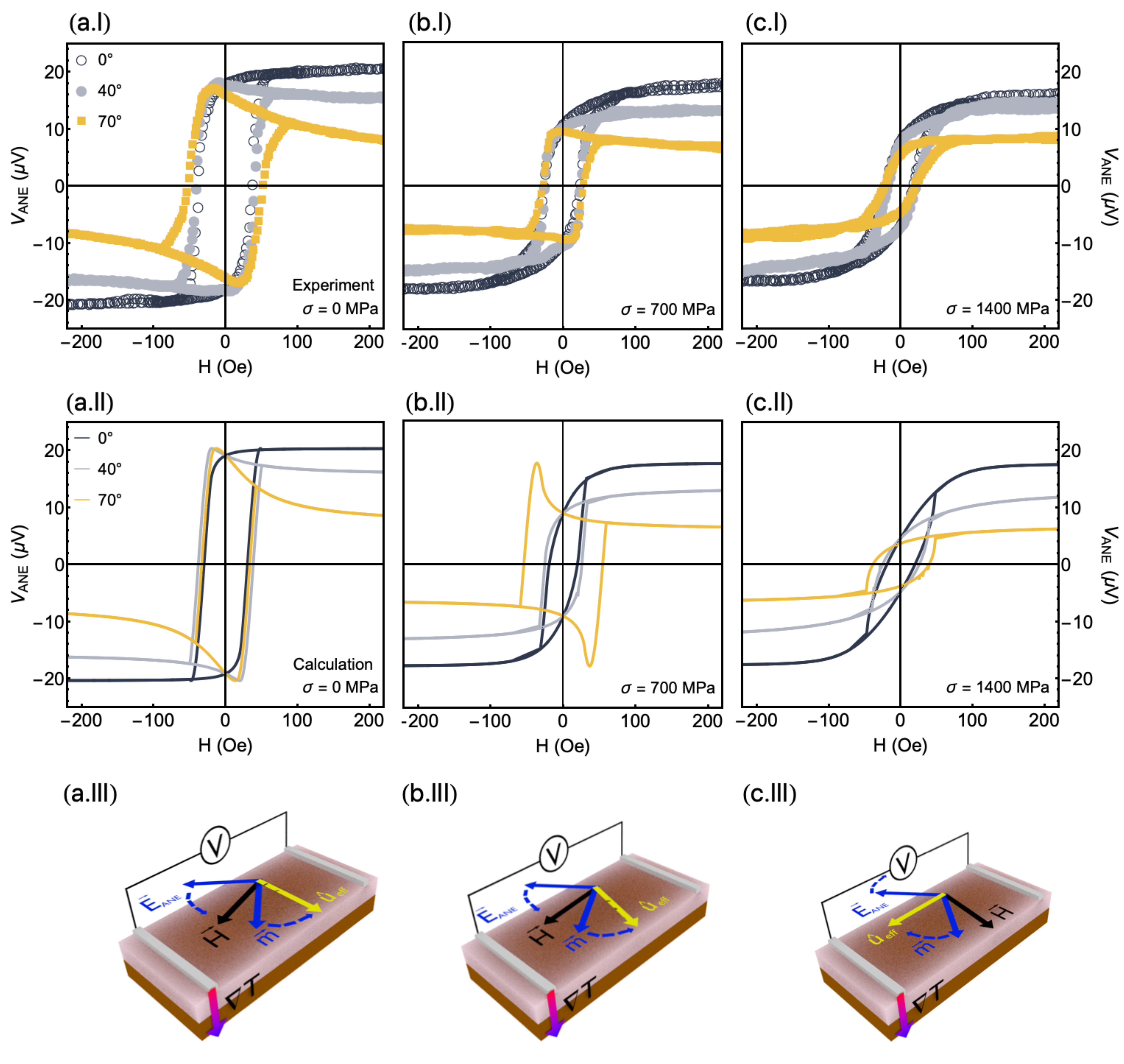} \vspace{-.1cm} 
\caption{{\bf Evolution of the thermoelectric voltage with the magnetic field and stress.} 
({\bf a}) Experimental results and numerical calculations of the $V_{ANE}$ response as a function of the magnetic field, for selected $\varphi_H$ values, for the flexible CoFeB film without stress. 
Below, illustration showing orientations of $\vec {E}_{ANE}$, $\vec H$, $\vec m$ and $\hat u_{eff}$. The dashed lines represent the direction in which $\vec m$ and $\vec {E}_{ANE}$ rotates with decreasing the magnetic field. 
Similar plots for the flexible CoFeB film submitted to stress values of ({\bf b}) $\sigma = 700$~MPa and ({\bf c}) $\sigma = 1400$~MPa. 
We set $\Delta T = 27$~K for the $V_{ANE}$ acquisition. 
For the numerical calculations, we consider the following parameters: $m_s= 625$~emu/cm$^3$, $\lambda_{s} = +30.0\times 10^{-6}$, $H_k= 42$~Oe, $\theta_k= 90^{\circ}$, $\varphi_k = 30^{\circ}$, and $t_{f}= 300$~nm, with $\lambda_N = 16.7 \times 10^{-6}$~V/KT in (a), $\lambda_N = 14.6 \times 10^{-6}$~V/KT in (b), and $\lambda_N = 13.9 \times 10^{-6}$~V/KT in (c). See the movie in Supplementary Information. 
}
\label{Fig_05}
\end{figure}

The most striking experimental and theoretical findings here are shown in Fig.~\ref{Fig_05}, which discloses the evolution of the $V_{ANE}$ response, as a function of the magnetic field, with the $\varphi_H$ and stress $\sigma$ values for the CoFeB film grown onto a flexible substrate. 
The numerical calculations for a film with uniaxial magnetic anisotropy and under stress are performed considering system parameters similar to those previously employed in Fig.~\ref{Fig_02}, except for the $\varphi_k$ and $\lambda_N$ values.
In this case, $m_s= 625$~emu/cm$^3$ and $\lambda_{s} = +30.0\times 10^{-6}$, $H_k= 42$~Oe, $\theta_k= 90^{\circ}$, $\varphi_k= 30^{\circ}$ and $t_{f}= 300$~nm. 
The orientation of $\hat u_k$, set by $\varphi_k$, was changed in order to mimic the magnetic properties of the flexible film. 
Further, we consider $\lambda_N = 16.7 \times 10^{-6}$, $14.6 \times 10^{-6}$ and $13.9 \times 10^{-6}$~V/KT (see Methods for details on the $\lambda_N$ estimation) for the calculations with $\sigma = 0$, $700$, and $1400$~MPa, respectively. 
In particular, the decrease in the $\lambda_N$ with $\sigma$ is straightly verified through the reduction of the $V_{ANE}^{Smax}$; and we associate it to modifications in the energy relaxation time close of the Fermi level, which can be altered by the stress application~\cite{PRB59pR9019}. 
Then, notice the striking agreement between our experiments and numerical calculations, including three important features: the own shape of the $V_{ANE}$ curves, the $\varphi_H$ dependence of the evolution in the shape of the curves, and amplitude of the $V_{ANE}$ response at high magnetic fields, i.e.~at the magnetic saturation state. 

From Fig.~\ref{Fig_05}(a), the $V_{ANE}$ curves for the film without stress exhibit all the features found for a film with uniaxial magnetic anisotropy, as expected; hence, their interpretation is the very same to that previously reported in Figs.~\ref{Fig_02}(b) and \ref{Fig_03}(a.II,b.II). 
With respect to the shape of the curves and its evolution with the orientation of the magnetic field, it is worth observing that the $V_{ANE}$ response pattern changes with $\varphi_H$ crossing over through $\varphi_{k_{eff}}$. 
Specifically, keeping in mind that in this case the effective magnetic anisotropy $\hat u_{k_{eff}}$ is the own uniaxial magnetic anisotropy $\hat u_k$ with $\varphi_{k} \approx 30^{\circ}$, the $V_{ANE}$ curves for $\varphi_H < \varphi_{k_{eff}}$ have shape mirroring the corresponding magnetization loops presented in Fig.~\ref{Fig_04}(a). 
At small $\varphi_H$ values, represented in Fig.~\ref{Fig_05}(a) by the curve for $\varphi_H = 0^\circ$, the magnetization is primarily kept close to the magnetic field direction and/or to the easy magnetization axis, thus favouring the alignment between $\vec E_{ANE}$ and the detection direction defined by the electrical contacts, and therefore $V_{ANE}$ is directly proportional to the magnitude of the magnetization. 
The $V_{ANE}$ response for $\varphi_H > \varphi_{k_{eff}}$ in turn, depicted by $\varphi_H = 40^\circ$ and $70^\circ$, discloses curves with completely different signatures, losing the squared shape. 
For these cases, at high magnetic field values, the magnetization is aligned with the field and, consequently, $V_{ANE}$ is drastically reduced; at the low magnetic field range, the magnetization $\vec m$ rotates and remains close to the easy magnetization axis $\hat {u}_{eff}$, as we can see in the schematic representation in Fig.~\ref{Fig_05}(a.III), leading to higher $V_{ANE}$ values. 

For the film under $\sigma \approx 700$~MPa, Fig.~\ref{Fig_05}(b) shows curves with a slightly different profile. 
In this case, the effective magnetic anisotropy is described by an angle $\varphi_{k_{eff}}$ that lies between $40^\circ$ and $70^\circ$, in a sense that the change in the $V_{ANE}$ response pattern takes place within this angular range. 
Hence, observe that the $V_{ANE}$ responses at $\varphi_H = 0^\circ$ and $40^\circ$ have the features of a magnetization curve, just presenting a decrease in the amplitude of the $V_{ANE}$ signal at the magnetic saturation state, as expected due to the thermoelectric voltage configuration in the experiment. 
At $\varphi_H = 70^\circ$, the $V_{ANE}$ shape changes considerably, suggesting the competition between the Zeeman interaction and the effective magnetic anisotropy. 
In this case, at high fields, the magnetization follows the magnetic field, leading to a decrease of $V_{ANE}$. 
However, as the field decreases, the Zeeman interaction is reduced and the magnetization $\vec m$ turns to $\hat{u}_{eff}$. 
As a consequence, there is the increase in the component of $\vec{E}_{ANE}$ along the $V_{ANE}$ detection direction, as we can see in the schematic representation in Fig.~\ref{Fig_05}(b.III). 
Obviously, as aforementioned, the discrepancy between experiment and theory at this low-field values is primarily associated magnetization process and to the existence of magnetic domains in the film, which is not taken into account in our theoretical approach as already discussed. 

At last, for the film under $\sigma \approx 1400$~MPa shown in Fig.~\ref{Fig_05}(c), the effective magnetic anisotropy resides in an angle $\varphi_{k_{eff}}> 70^\circ$, specifically close to $\varphi_{k_{eff}}\approx 90^\circ$.
As a result, the presented $V_{ANE}$ curves have the very same features for all the selected $\varphi_H$ values. 
At $\varphi_H = 0^\circ$, the $V_{ANE}$ curve mirrors the corresponding magnetization loop shown in Fig.~\ref{Fig_04}(a). 
As $\varphi_H$ increases, we find the decrease in the amplitude of the $V_{ANE}$ signal at the magnetic saturation state, as expected. 
Once the magnetic field decreases, irrespective of $\varphi_H$, the magnetization $\vec m$ rotates to the orientation $\hat{u}_{eff}$ of the effective magnetic anisotropy axis.
Given that $\vec{E}_{ANE}$ tends to be almost transverse to the $V_{ANE}$ detection direction, as we can confirm through the schematic representation in Fig.~\ref{Fig_05}(c.III), a decrease in the $V_{ANE}$ value is also found at low field values and, as a consequence, the evolution in the shape on the $V_{ANE}$ curves is not observed here. 
In particular, for $\varphi_H = 90^\circ$, $V_{ANE}\approx 0$ for the whole magnetic field range. 

\begin{figure}[!t]\centering
\vspace{.5cm}
\includegraphics[width=10.5cm]{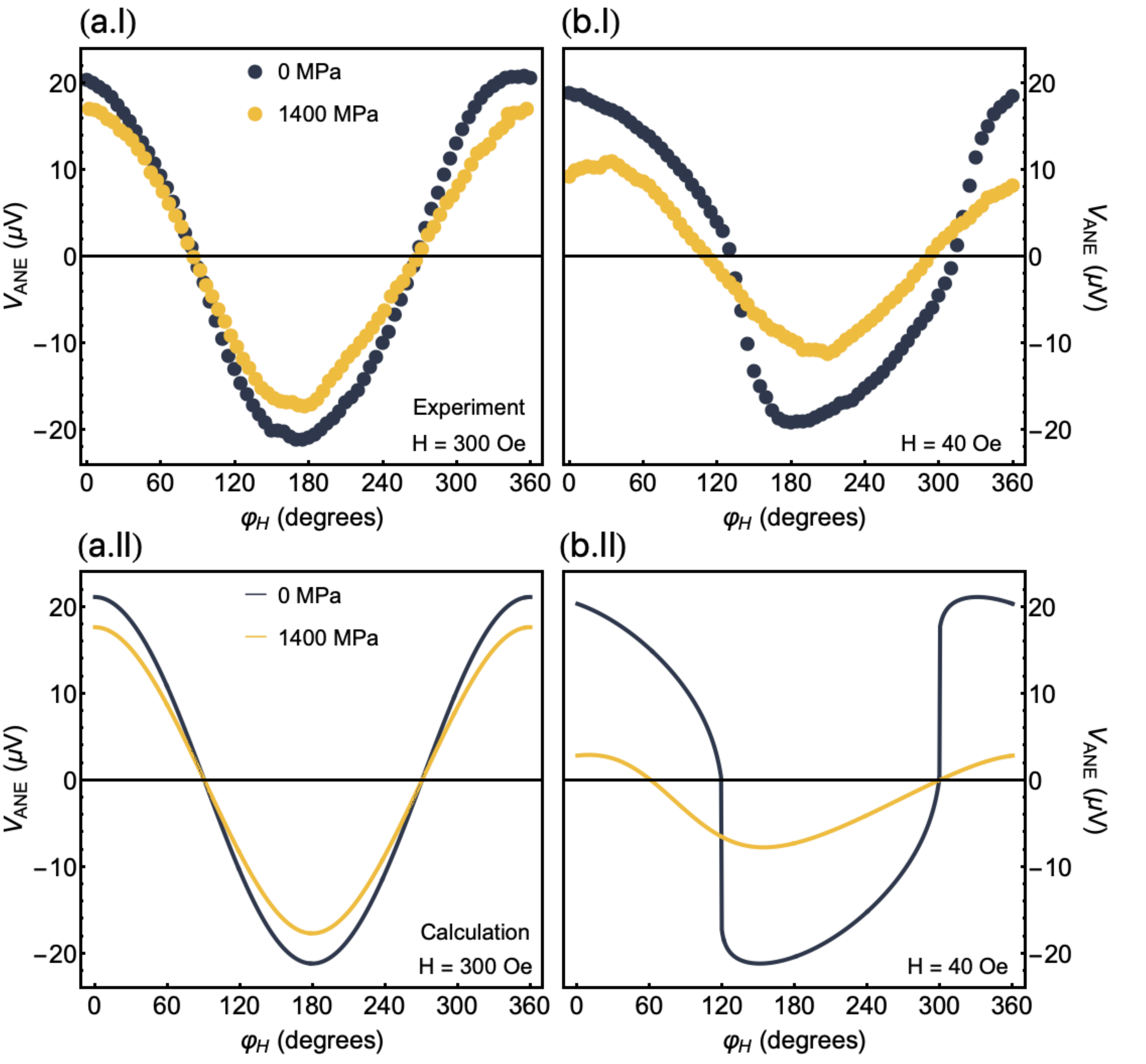} \vspace{-.1cm} 
\caption{{\bf Angular dependence of the thermoelectric voltage.} 
({\bf a}) Experimental results and numerical calculations for the $V_{ANE}$, at $H=+ 300$~Oe and with $\Delta T = 27$~K, as a function of $\varphi_H$ for the CoFeB film under selected stress levels. 
At this field value, the sample is magnetically saturated. 
({\bf b}) Similar plot for the $V_{ANE}$ at $H= + 40$~Oe, where the sample is an unsaturated state.
For the numerical calculations, we consider the very same parameters emloyed in Fig.~\ref{Fig_05}, i.e.\ $m_s= 625$~emu/cm$^3$, $\lambda_{s} = +30.0\times 10^{-6}$, $H_k= 42$~Oe, $\theta_k= 90^{\circ}$, $\varphi_k = 30^{\circ}$ and $t_{f}= 300$~nm, with $\lambda_{N}=16.7\times10^{-6}$~V/KT when $\sigma = 0$~MPa, and $\lambda_{N}=13.9\times10^{-6}$~V/KT when $\sigma = 1400$~MPa. 
}
\label{Fig_06}
\end{figure}

Finally, looking at the angular dependence of the $V_{ANE}$ signal amplitude at a given magnetic field, Fig.~\ref{Fig_06} shows the thermoelectric voltage as a function of $\varphi_H$ for the CoFeB film under selected stress levels. 
At $H=300$~Oe, Fig~\ref{Fig_06}(a), the sample is magnetically saturated, irrespective on the stress level. 
As a consequence, the $V_{ANE}$ response is precisely the one described by Eq.~(\ref{Fig_05}), with a well-defined cosine shape. 
Another important test of consistency of our approach is also given by the $V_{ANE}$ behavior at unsaturated states. 
The curves are a result of the competition between the uniaxial magnetic anisotropy and the magnetoelastic anisotropy, leading to considerable changes in the shape and amplitude of the $V_{ANE}$ response. 
Notice the striking quantitative agreement between experiment and theory in Fig~\ref{Fig_06}(b). 

Hence, we are able to describe through numerical calculations all the main features of the $V_{ANE}$ response. Thus, we provide experimental evidence to confirm the validity of the theoretical approach to describe the magnetic properties and anomalous Nernst effect in ferromagnetic magnetostrictive films having uniaxial magnetic anisotropy and submitted to external stress.

\section*{Discussion}

In summary, we have performed a theoretical and experimental investigation of the magnetic properties and anomalous Nernst effect in a flexible magnetostrictive film with induced uniaxial magnetic anisotropy and under external stress. 
Our findings raise numerous interesting issues on the anomalous Nernst effect in nanostructured magnetic materials. 
In particular, they show how the magnetization behavior and the thermoelectric voltage response evolve with both, the magnetic field and external stress. 
Hence, by comparing our experiments with numerical calculations, we elucidate the magnetic properties and thermoelectric voltage and demonstrate the possibility of tailoring the anomalous Nernst effect in a flexible magnetostrictive film. 
The quantitative agreement between experiment and numerical calculations provides evidence to confirm the validity of the theoretical approach to describe the magnetic properties and anomalous Nernst effect in ferromagnetic magnetostrictive films having uniaxial magnetic anisotropy and submitted to external stress. 
The results place flexible magnetostrictive systems as promising candidate for active elements in functionalized touch electronic devices.

\section*{Methods}

\noindent{\bf Estimation of $\Delta T_{f}$. } 
The relation between the experimentally measured temperature variation $\Delta T$ across the sample and the effective temperature variation $\Delta T_f$ in the film is given by 
\begin{equation}
\Delta T_{f} =\frac{t_{f} K_{sub}}{t_{sub}K_{f}} \Delta T,
\label{dtf}
\end{equation}
\noindent where $K_{sub}$ and $t_{sub}$ are the thermal conductivity and thickness of the substrate, while $K_f$ and $t_{f}$ are the respective quantities for the ferromagnetic film. 
Here, we consider $K_{sub} = 0.12$~W/Km and $t_{sub}=0.15$~mm for the flexible substrate (Kapton$^{\textrm{\tiny \textregistered}}$), while $K_f=86.7$~W/Km~\cite{SR5P10249} and $t_{f}= 300$~nm for our CoFeB film. 

\begin{figure}[!h]\centering
\vspace{.5cm}
\includegraphics[width=5.5cm]{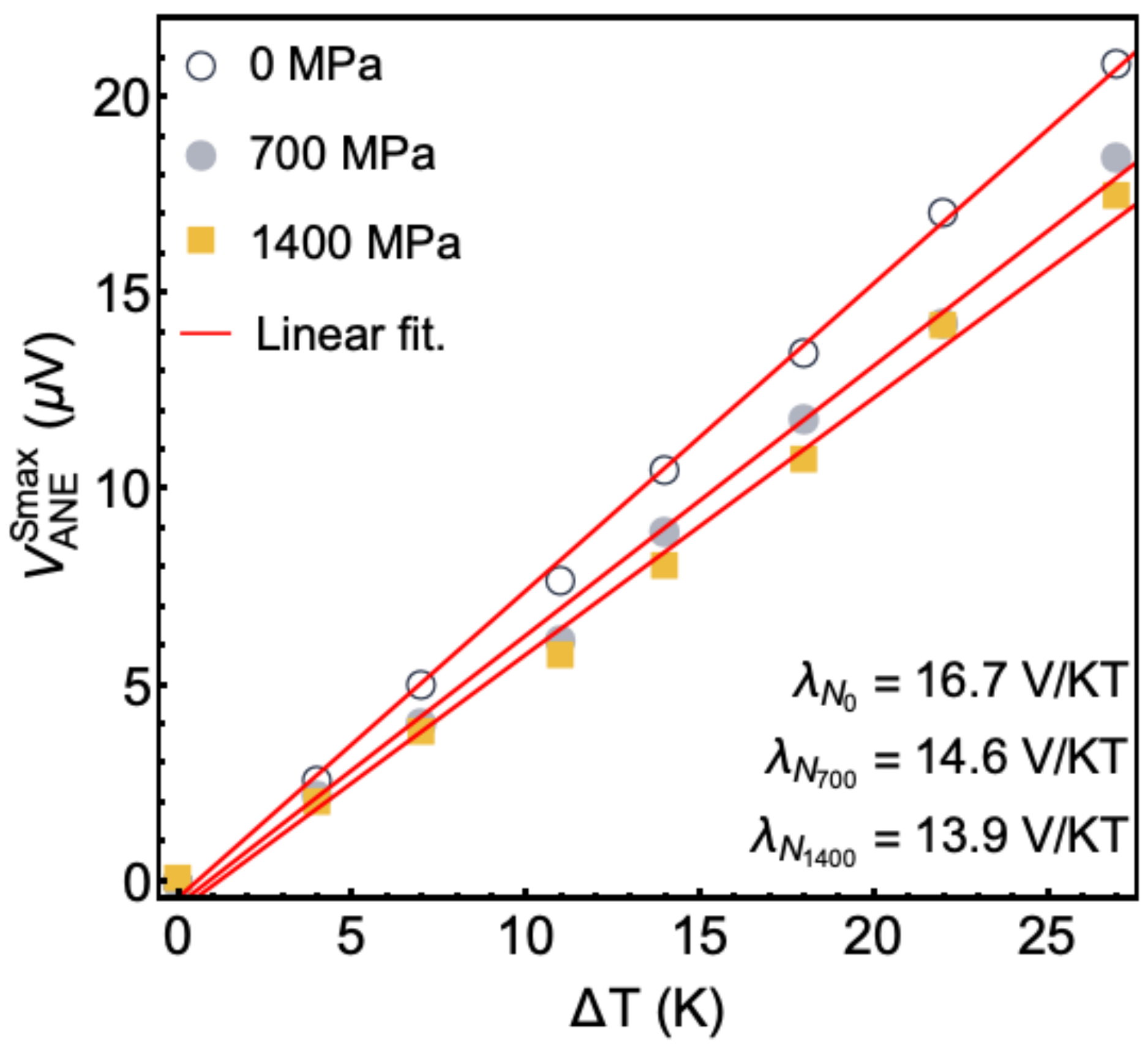} \vspace{-.1cm} 
\caption{{\bf Estimation of the Nernst coefficient $\lambda_N$.} 
$V_{ANE}^{Smax}$ as a function of $\Delta T$ for distinct applied stress $\sigma$. The $\lambda_N$ is obtained by using Eqs.~(\ref{dtf}) and (\ref{Nernstcoef}). }
\label{Fig_07}
\end{figure}

\noindent{\bf Estimation of $\lambda_N$. } 
We estimate the $\lambda_N$ values for each applied stress $\sigma$ level. 
This coefficient is obtained from the linear fitting of the experimental data of $V_{ANE}^{Smax}$ as a function of $\Delta T$, as shown in Fig.~\ref{Fig_07}. 
In particular, $\Delta T_{f}$ is calculated by using the Eq.~(\ref{dtf}) and thus $\lambda_N$ is given by using Eq.~(\ref{Nernstcoef}). 
As a result, we assume $\lambda_N = 16.7 \times 10^{-6}$, $14.6 \times 10^{-6}$ and $13.9 \times 10^{-6}$~V/KT for the calculations with $\sigma = 0$, $700$, and $1400$~MPa, respectively. 

\noindent{\bf Sample preparation.} 
We investigate the anomalous Nernst effect in Co$_{40}$Fe$_{40}$B$_{20}$ (CoFeB) films with thickness of $300$ nm grown onto rigid (glass) and flexible (Kapton$^{\textrm{\tiny \textregistered}}$) substrates. 
The films are deposited by magnetron sputtering onto substrates with dimensions of $10 \times 6$ mm$^2$.
The deposition process is carried out with the following parameters: base pressure of $5 \times 10^{-6}$ Torr, deposition pressure of $3.0 \times 10^{-3}$ torr with a $99.99\%$ pure Ar at $50$~sccm constant flow, and with $50$~W set in the DC source. 
By using these parameters, the deposition rate is $1.86$~nm/s.
During the deposition, a constant in-plane magnetic field of $1$~kOe is applied perpendicular to the main axis of the substrate in order to induce an uniaxial magnetic anisotropy.

\noindent{\bf Magnetization measurements.} 
Quasi-static magnetization curves are obtained within the range between $\varphi_H=0^\circ$ (perpendicular) and $\varphi_H=90^\circ$ (along the main axis of the films), in order to verify the magnetic properties.
The curves are acquired at room temperature using a Lake Shore model 7404 vibrating sample magnetometer, with maximum in-plane magnetic field of $\pm 300$ Oe. 

\noindent{\bf Thermoelectric voltage experiments.} 
We employ a homemade experimental system to measure the thermoelectric voltage associated with the anomalous Nernst effect.
In our experimental setup, the temperature gradient $\nabla T$ is normal to the film plane. 
A Peltier module is used to heat or cool the top of the sample while the substrate is kept in thermal contact with a copper block at room temperature. 
The temperature difference $\Delta T$ across the sample is measured with a differential thermocouple.
The magnetic field is kept in the film plane, although its orientation can be modified by varying $\varphi_H$ from $0^\circ$ up to $360^\circ$, having as reference the dashed line indicated in Fig.~\ref{Fig_01}(b).
Finally, the $V_{ANE}$ detection is performed using a nanovoltmeter with electrical contacts in the ends of the main axis of the film, also illustrated in Fig.~\ref{Fig_01}(a,b), whose separation distance is $L=5.0$~mm. 

\noindent{\bf Stress application.} 
The magnetization measurements and thermoelectric voltage experiments are carried out in CoFeB films grown onto both, rigid and flexible substrates. 
Considering the flexible films, we perform acquisitions for the sample with and without stress. 
The stress is applied in the film by bending the sample, along the main axis, during the measurement. 
The magnitude of the stress $\sigma$ is calculated by using the procedure previously employed in Refs.~\cite{JMMM420p81, JMMM453p30}. 
By bending the flexible substrate, we are able to induce tensile stress in the film, thus modifying the effective magnetic anisotropy~\cite{APL100p122407, APL106p162405, JMMM420p81}.

\section*{Acknowledgements}
~The authors acknowledge financial support from the Brazilian agencies CNPq and CAPES.

\section*{Author contributions statement}
~M.A.C. and F.B were responsible for the theoretical approach and performed the numerical calculations. 
A.B.O. and C.C. were responsible for the development of the experimental system. 
A.S.M. and R.D.D.P produced the samples and performed the experimental measurements. 
All authors contributed to interpreting the results, and write and improve the text of the manuscript.

\section*{Competing Financial Interests statement}
~The authors declare no competing interests.  

\section*{Additional information}
~Correspondence and requests for materials shall be addressed to M.A.C.


\begin{thebibliography}{10}
\expandafter\ifx\csname url\endcsname\relax
  \def\url#1{\texttt{#1}}\fi
\expandafter\ifx\csname urlprefix\endcsname\relax\def\urlprefix{URL }\fi
\expandafter\ifx\csname doiprefix\endcsname\relax\def\doiprefix{DOI }\fi
\providecommand{\bibinfo}[2]{#2}
\providecommand{\eprint}[2][]{\url{#2}}

\bibitem{SR6p23114}
\bibinfo{author}{Kirihara, A.} \emph{et~al.}
\newblock \bibinfo{journal}{\bibinfo{title}{{Flexible heat-flow sensing sheets
  based on the longitudinal spin Seebeck effect using one-dimensional
  spin-current conducting films}}}.
\newblock {\emph{\JournalTitle{Sci. Rep.}}} \textbf{\bibinfo{volume}{6}},
  \bibinfo{pages}{23114} (\bibinfo{year}{2016}).

\bibitem{JMMM290p795}
\bibinfo{author}{Dokupil, S.}, \bibinfo{author}{Bootsmann, M.~T.},
  \bibinfo{author}{Stein, S.}, \bibinfo{author}{L{\"{o}}hndorf, M.} \&
  \bibinfo{author}{Quandt, E.}
\newblock \bibinfo{journal}{\bibinfo{title}{Positive/negative magnetostrictive
  {GMR} trilayer systems as strain gauges}}.
\newblock {\emph{\JournalTitle{J. Magn. Magn. Mater.}}}
  \textbf{\bibinfo{volume}{290}}, \bibinfo{pages}{795--799}
  (\bibinfo{year}{2005}).

\bibitem{EES7p885910}
\bibinfo{author}{Boona, S.~R.}, \bibinfo{author}{Myers, R.~C.} \&
  \bibinfo{author}{Heremans, J.~P.}
\newblock \bibinfo{journal}{\bibinfo{title}{{Spin caloritronics}}}.
\newblock {\emph{\JournalTitle{Energy and Environmental Science}}}
  \textbf{\bibinfo{volume}{7}}, \bibinfo{pages}{885--910}
  (\bibinfo{year}{2014}).

\bibitem{NMAT11p391399}
\bibinfo{author}{Bauer, G. E.~W.}, \bibinfo{author}{Saitoh, E.} \&
  \bibinfo{author}{Wees, B. J.~V.}
\newblock \bibinfo{journal}{\bibinfo{title}{{Spin caloritronics}}}.
\newblock {\emph{\JournalTitle{Nat. Mater.}}} \textbf{\bibinfo{volume}{11}},
  \bibinfo{pages}{391--399} (\bibinfo{year}{2012}).

\bibitem{PRL108p106602}
\bibinfo{author}{Weiler, M.} \emph{et~al.}
\newblock \bibinfo{journal}{\bibinfo{title}{{Local Charge and Spin Currents in
  Magnetothermal Landscapes}}}.
\newblock {\emph{\JournalTitle{Phys. Rev. Lett.}}}
  \textbf{\bibinfo{volume}{108}}, \bibinfo{pages}{106602}
  (\bibinfo{year}{2012}).

\bibitem{APL106p212407}
\bibinfo{author}{Tian, D.}, \bibinfo{author}{Li, Y.}, \bibinfo{author}{Qu, D.},
  \bibinfo{author}{Jin, X.} \& \bibinfo{author}{Chien, C.~L.}
\newblock \bibinfo{journal}{\bibinfo{title}{{Separation of spin Seebeck effect
  and anomalous Nernst effect in Co/Cu/YIG}}}.
\newblock {\emph{\JournalTitle{App. Phys. Lett.}}}
  \textbf{\bibinfo{volume}{106}}, \bibinfo{pages}{212407}
  (\bibinfo{year}{2015}).

\bibitem{SREP7p6165}
\bibinfo{author}{Kannan, H.}, \bibinfo{author}{X.~Fan, H.}, \bibinfo{author}{X,
  C.}, \bibinfo{author}{Han} \& \bibinfo{author}{Xiao, J.~Q.}
\newblock \bibinfo{journal}{\bibinfo{title}{Thickness dependence of anomalous
  {N}ernst coefficient and longitudinal spin seebeck effect in ferromagnetic
  {Ni$_{x}$Fe$_{100-x}$} films}}.
\newblock {\emph{\JournalTitle{Sci. Rep.}}} \textbf{\bibinfo{volume}{7}},
  \bibinfo{pages}{6165} (\bibinfo{year}{2017}).

\bibitem{APE10p073005}
\bibinfo{author}{Isogami, S.}, \bibinfo{author}{Takanashi, K.} \&
  \bibinfo{author}{Mizuguchi, M.}
\newblock \bibinfo{journal}{\bibinfo{title}{{Dependence of anomalous Nernst
  effect on crystal orientation in highly ordered $\gamma$-Fe$_4$N films with
  anti-perovskite structure}}}.
\newblock {\emph{\JournalTitle{Applied Physics Express}}}
  \textbf{\bibinfo{volume}{10}}, \bibinfo{pages}{73005} (\bibinfo{year}{2017}).

\bibitem{PRB96p174406}
\bibinfo{author}{Chuang, T.~C.}, \bibinfo{author}{Su, P.~L.},
  \bibinfo{author}{Wu, P.~H.} \& \bibinfo{author}{Huang, S.~Y.}
\newblock \bibinfo{journal}{\bibinfo{title}{{Enhancement of the anomalous
  Nernst effect in ferromagnetic thin films}}}.
\newblock {\emph{\JournalTitle{Phys. Rev. B}}} \textbf{\bibinfo{volume}{96}},
  \bibinfo{pages}{174406} (\bibinfo{year}{2017}).

\bibitem{APL106p252405}
\bibinfo{author}{Hasegawa, K.} \emph{et~al.}
\newblock \bibinfo{journal}{\bibinfo{title}{{Material dependence of anomalous
  Nernst effect in perpendicularly magnetized ordered-alloy thin films}}}.
\newblock {\emph{\JournalTitle{App. Phys. Lett.}}}
  \textbf{\bibinfo{volume}{106}}, \bibinfo{pages}{252405}
  (\bibinfo{year}{2015}).

\bibitem{APL105p103504}
\bibinfo{author}{Tang, Z.} \emph{et~al.}
\newblock \bibinfo{journal}{\bibinfo{title}{{Magneto-mechanical coupling effect
  in amorphous Co$_{40}$Fe$_{40}$B$_{20}$ films grown on flexible
  substrates}}}.
\newblock {\emph{\JournalTitle{App. Phys. Lett.}}}
  \textbf{\bibinfo{volume}{105}}, \bibinfo{pages}{103504}
  (\bibinfo{year}{2014}).

\bibitem{JAP113p213909}
\bibinfo{author}{Conca, A.} \emph{et~al.}
\newblock \bibinfo{journal}{\bibinfo{title}{{Low spin-wave damping in amorphous
  Co$_{40}$Fe$_{40}$B$_{20}$ thin films}}}.
\newblock {\emph{\JournalTitle{J. Appl. Phys.}}}
  \textbf{\bibinfo{volume}{113}}, \bibinfo{pages}{213909}
  (\bibinfo{year}{2013}).

\bibitem{JAP100p053903}
\bibinfo{author}{Bilzer, C.} \emph{et~al.}
\newblock \bibinfo{journal}{\bibinfo{title}{{Study of the dynamic magnetic
  properties of soft CoFeB films}}}.
\newblock {\emph{\JournalTitle{J. Appl. Phys.}}}
  \textbf{\bibinfo{volume}{100}}, \bibinfo{pages}{53903}
  (\bibinfo{year}{2006}).

\bibitem{JAP110p033910}
\bibinfo{author}{Liu, X.}, \bibinfo{author}{Zhang, W.},
  \bibinfo{author}{Carter, M.~J.} \& \bibinfo{author}{Xiao, G.}
\newblock \bibinfo{journal}{\bibinfo{title}{{Ferromagnetic resonance and
  damping properties of CoFeB thin films as free layers in MgO-based magnetic
  tunnel junctions}}}.
\newblock {\emph{\JournalTitle{J. Appl. Phys.}}}
  \textbf{\bibinfo{volume}{110}}, \bibinfo{pages}{33910}
  (\bibinfo{year}{2011}).

\bibitem{NM3p862}
\bibinfo{author}{Parkin, S. S.~P.} \emph{et~al.}
\newblock \bibinfo{journal}{\bibinfo{title}{{Giant tunnelling magnetoresistance
  at room temperature with MgO (100) tunnel barriers}}}.
\newblock {\emph{\JournalTitle{Nat. Mater.}}} \textbf{\bibinfo{volume}{3}},
  \bibinfo{pages}{862--867} (\bibinfo{year}{2004}).

\bibitem{JAP119p133903}
\bibinfo{author}{Jamali, M.}, \bibinfo{author}{Klemm~Smith, A.} \&
  \bibinfo{author}{Wang, J.-P.}
\newblock \bibinfo{journal}{\bibinfo{title}{{Nonreciprocal behavior of the spin
  pumping in ultra-thin film of CoFeB}}}.
\newblock {\emph{\JournalTitle{J. Appl. Phys.}}}
  \textbf{\bibinfo{volume}{119}}, \bibinfo{pages}{133903}
  (\bibinfo{year}{2016}).

\bibitem{JAP117p163901}
\bibinfo{author}{Ruiz-Calaforra, A.} \emph{et~al.}
\newblock \bibinfo{journal}{\bibinfo{title}{{The role of the non-magnetic
  material in spin pumping and magnetization dynamics in NiFe and CoFeB
  multilayer systems}}}.
\newblock {\emph{\JournalTitle{J. Appl. Phys.}}}
  \textbf{\bibinfo{volume}{117}}, \bibinfo{pages}{163901}
  (\bibinfo{year}{2015}).

\bibitem{JAP111p07C520}
\bibinfo{author}{Liebing, N.} \emph{et~al.}
\newblock \bibinfo{journal}{\bibinfo{title}{{Determination of spin-dependent
  Seebeck coefficients of CoFeB/MgO/CoFeB magnetic tunnel junction
  nanopillars}}}.
\newblock {\emph{\JournalTitle{J. Appl. Phys.}}}
  \textbf{\bibinfo{volume}{111}}, \bibinfo{pages}{07C520}
  (\bibinfo{year}{2012}).

\bibitem{PRB88p064403}
\bibinfo{author}{Rojas-S{\'{a}}nchez, J.-C.} \emph{et~al.}
\newblock \bibinfo{journal}{\bibinfo{title}{{Spin pumping and inverse spin Hall
  effect in germanium}}}.
\newblock {\emph{\JournalTitle{Phys. Rev. B}}} \textbf{\bibinfo{volume}{88}},
  \bibinfo{pages}{64403} (\bibinfo{year}{2013}).

\bibitem{NMAT9p721724}
\bibinfo{author}{Ikeda, S.} \emph{et~al.}
\newblock \bibinfo{journal}{\bibinfo{title}{A perpendicular-anisotropy
  {C}o{F}e{B}-{M}g{O} magnetic tunnel junction}}.
\newblock {\emph{\JournalTitle{Nat. Mater.}}} \textbf{\bibinfo{volume}{9}},
  \bibinfo{pages}{721--729} (\bibinfo{year}{2010}).

\bibitem{JMMM462p2940}
\bibinfo{author}{Gayen, A.}, \bibinfo{author}{Prasad, G.~K.},
  \bibinfo{author}{Umadevi, K.}, \bibinfo{author}{Chelvane, J.~A.} \&
  \bibinfo{author}{Alagarsamy, P.}
\newblock \bibinfo{journal}{\bibinfo{title}{{Interlayer coupling in symmetric
  and asymmetric CoFeB based trilayer films with different domain structures:
  Role of spacer layer and temperature}}}.
\newblock {\emph{\JournalTitle{Journal of Magnetism and Magnetic Materials}}}
  \textbf{\bibinfo{volume}{462}}, \bibinfo{pages}{29--40}
  (\bibinfo{year}{2018}).

\bibitem{PRB83p212404}
\bibinfo{author}{Hindmarch, A.~T.}, \bibinfo{author}{Rushforth, A.~W.},
  \bibinfo{author}{Campion, R.~P.}, \bibinfo{author}{Marrows, C.~H.} \&
  \bibinfo{author}{Gallagher, B.~L.}
\newblock \bibinfo{journal}{\bibinfo{title}{{Origin of in-plane uniaxial
  magnetic anisotropy in CoFeB amorphous ferromagnetic thin films}}}.
\newblock {\emph{\JournalTitle{Phys. Rev. B}}} \textbf{\bibinfo{volume}{83}},
  \bibinfo{pages}{212404} (\bibinfo{year}{2011}).

\bibitem{JMMM426p444}
\bibinfo{author}{Tang, Z.} \emph{et~al.}
\newblock \bibinfo{journal}{\bibinfo{title}{Thickness dependence of magnetic
  anisotropy and domains in amorphous {C}o$_{40}${F}e$_{40}${B}$_{20}$ thin
  films grown on {PET} flexible substrates}}.
\newblock {\emph{\JournalTitle{J. Magn. Magn. Mater.}}}
  \textbf{\bibinfo{volume}{426}}, \bibinfo{pages}{444 -- 449}
  (\bibinfo{year}{2017}).

\bibitem{AFM26p4704}
\bibinfo{author}{Vemulkar, T.}, \bibinfo{author}{Mansell, R.},
  \bibinfo{author}{Fern{\'a}ndez-Pacheco, A.} \& \bibinfo{author}{Cowburn,
  R.~P.}
\newblock \bibinfo{journal}{\bibinfo{title}{Toward flexible spintronics:
  Perpendicularly magnetized synthetic antiferromagnetic thin films and
  nanowires on polyimide substrates}}.
\newblock {\emph{\JournalTitle{Adv. Funct. Mat.}}}
  \textbf{\bibinfo{volume}{26}}, \bibinfo{pages}{4704--4711}
  (\bibinfo{year}{2016}).

\bibitem{AIPA6p056106}
\bibinfo{author}{Qiao, X.} \emph{et~al.}
\newblock \bibinfo{journal}{\bibinfo{title}{{Tuning magnetic anisotropy of
  amorphous CoFeB film by depositing on convex flexible substrates}}}.
\newblock {\emph{\JournalTitle{AIP Advances}}} \textbf{\bibinfo{volume}{6}},
  \bibinfo{pages}{56106} (\bibinfo{year}{2016}).

\bibitem{NATC3p1259}
\bibinfo{author}{Yi, H.~T.}, \bibinfo{author}{Payne, M.~M.},
  \bibinfo{author}{Anthony, J.~E.} \& \bibinfo{author}{Podzorov, V.}
\newblock \bibinfo{journal}{\bibinfo{title}{{Ultra-flexible solution-processed
  organic field-effect transistors}}}.
\newblock {\emph{\JournalTitle{Nat. Commun.}}} \textbf{\bibinfo{volume}{3}},
  \bibinfo{pages}{1259} (\bibinfo{year}{2012}).

\bibitem{ADVM25p5997}
\bibinfo{author}{Hammock, M.~L.}, \bibinfo{author}{Chortos, A.},
  \bibinfo{author}{Tee, B. C.-K.}, \bibinfo{author}{Tok, J. B.-H.} \&
  \bibinfo{author}{Bao, Z.}
\newblock \bibinfo{journal}{\bibinfo{title}{25th anniversary article: The
  evolution of electronic skin (e-skin): A brief history, design
  considerations, and recent progress}}.
\newblock {\emph{\JournalTitle{Adv. Mater.}}} \textbf{\bibinfo{volume}{25}},
  \bibinfo{pages}{5997--6038} (\bibinfo{year}{2013}).

\bibitem{ADVM26p13361342}
\bibinfo{author}{Xuewen, W.}, \bibinfo{author}{Yang, G.},
  \bibinfo{author}{Zuoping, X.}, \bibinfo{author}{Zheng, C.} \&
  \bibinfo{author}{Ting, Z.}
\newblock \bibinfo{journal}{\bibinfo{title}{{Silk-Molded Flexible,
  Ultrasensitive, and Highly Stable Electronic Skin for Monitoring Human
  Physiological Signals}}}.
\newblock {\emph{\JournalTitle{Adv. Mater.}}} \textbf{\bibinfo{volume}{26}},
  \bibinfo{pages}{1336--1342} (\bibinfo{year}{2014}).

\bibitem{NAT428p911918}
\bibinfo{author}{Forrest, S.~R.}
\newblock \bibinfo{journal}{\bibinfo{title}{{The path to ubiquitous and
  low-cost organic electronic appliances on plastic}}}.
\newblock {\emph{\JournalTitle{Nature}}} \textbf{\bibinfo{volume}{428}},
  \bibinfo{pages}{911--918} (\bibinfo{year}{2004}).

\bibitem{JMMM453p30}
\bibinfo{author}{Correa, M.~A.} \& \bibinfo{author}{Bohn, F.}
\newblock \bibinfo{journal}{\bibinfo{title}{{Manipulating the magnetic
  anisotropy and magnetization dynamics by stress: Numerical calculation and
  experiment}}}.
\newblock {\emph{\JournalTitle{J. Magn. Magn. Mater.}}}
  \textbf{\bibinfo{volume}{453}}, \bibinfo{pages}{30--35}
  (\bibinfo{year}{2018}).

\bibitem{Cullity}
\bibinfo{author}{Melorose, J.}, \bibinfo{author}{Perroy, R.} \&
  \bibinfo{author}{Careas, S.}
\newblock \emph{\bibinfo{title}{{Introduction to Magnetic materials}}},
  vol.~\bibinfo{volume}{1} (\bibinfo{publisher}{Addison-Wesley},
  \bibinfo{address}{New York}, \bibinfo{year}{2015}).

\bibitem{JMMM420p81}
\bibinfo{author}{Agra, K.} \emph{et~al.}
\newblock \bibinfo{journal}{\bibinfo{title}{{Handling magnetic anisotropy and
  magnetoimpedance effect in flexible multilayers under external stress}}}.
\newblock {\emph{\JournalTitle{J. Magn. Magn. Mater.}}}
  \textbf{\bibinfo{volume}{420}}, \bibinfo{pages}{81--87}
  (\bibinfo{year}{2016}).

\bibitem{JAP109p07D736}
\bibinfo{author}{Naganuma, H.}, \bibinfo{author}{Oogane, M.} \&
  \bibinfo{author}{Ando, Y.}
\newblock \bibinfo{journal}{\bibinfo{title}{{Exchange biases of Co, Py,
  Co$_{40}$Fe$_{40}$B$_{20}$, Co$_{75}$Fe$_{25}$, and Co$_{50}$Fe$_{50}$ on
  epitaxial BiFeO$_{3}$ films prepared by chemical solution deposition}}}.
\newblock {\emph{\JournalTitle{J. Appl. Phys.}}}
  \textbf{\bibinfo{volume}{109}}, \bibinfo{pages}{07D736}
  (\bibinfo{year}{2011}).

\bibitem{JAP97p10C906}
\bibinfo{author}{Wang, D.}, \bibinfo{author}{Nordman, C.},
  \bibinfo{author}{Qian, Z.}, \bibinfo{author}{Daughton, J.~M.} \&
  \bibinfo{author}{Myers, J.}
\newblock \bibinfo{journal}{\bibinfo{title}{{Magnetostriction effect of
  amorphous CoFeB thin films and application in spin- dependent tunnel
  junctions Magnetostriction effect of amorphous CoFeB thin films and
  application in spin-dependent tunnel junctions}}}.
\newblock {\emph{\JournalTitle{J. Appl. Phys.}}} \textbf{\bibinfo{volume}{97}},
  \bibinfo{pages}{10C906} (\bibinfo{year}{2005}).

\bibitem{TSF520p2173}
\bibinfo{author}{Marques, M.~S.} \emph{et~al.}
\newblock \bibinfo{journal}{\bibinfo{title}{{High frequency magnetic behavior
  through the magnetoimpedance effect in CoFeB/(Ta, Ag, Cu) multilayered
  ferromagnetic thin films}}}.
\newblock {\emph{\JournalTitle{Thin Solid Films}}}
  \textbf{\bibinfo{volume}{520}}, \bibinfo{pages}{2173--2177}
  (\bibinfo{year}{2012}).

\bibitem{PTRS240p599}
\bibinfo{author}{Stoner, E.} \& \bibinfo{author}{Wohlfarth, E.}
\newblock \bibinfo{journal}{\bibinfo{title}{{A mechanism of magnetic hysteresis
  in heteresis in heterogeneous alloys}}}.
\newblock {\emph{\JournalTitle{Phil. Trans. Roy.Soc.}}}
  \textbf{\bibinfo{volume}{240}}, \bibinfo{pages}{599 -- 642}
  (\bibinfo{year}{1948}).

\bibitem{PRB59pR9019}
\bibinfo{author}{Suryanarayanan, R.}, \bibinfo{author}{Gasumyants, V.} \&
  \bibinfo{author}{Ageev, N.}
\newblock \bibinfo{journal}{\bibinfo{title}{Anomalous nernst effect in
  {La}$_{0.88}${Mn}{O}$_{3}$}}.
\newblock {\emph{\JournalTitle{Phys. Rev. B}}} \textbf{\bibinfo{volume}{59}},
  \bibinfo{pages}{R9019--R9022} (\bibinfo{year}{1999}).

\bibitem{SR5P10249}
\bibinfo{author}{Lee, K.~D.} \emph{et~al.}
\newblock \bibinfo{journal}{\bibinfo{title}{{Thermoelectric signal enhancement
  by reconciling the spin seebeck and anomalous nernst effects in
  ferromagnet/non-magnet multilayers}}}.
\newblock {\emph{\JournalTitle{Sci. Rep.}}} \textbf{\bibinfo{volume}{5}},
  \bibinfo{pages}{10249} (\bibinfo{year}{2015}).

\bibitem{APL100p122407}
\bibinfo{author}{Dai, G.} \emph{et~al.}
\newblock \bibinfo{journal}{\bibinfo{title}{{Mechanically tunable magnetic
  properties of Fe$_{81}$Ga$_{19}$ films grown on flexible substrates}}}.
\newblock {\emph{\JournalTitle{App. Phys. Lett.}}}
  \textbf{\bibinfo{volume}{100}}, \bibinfo{pages}{122407}
  (\bibinfo{year}{2012}).

\bibitem{APL106p162405}
\bibinfo{author}{Yu, Y.} \emph{et~al.}
\newblock \bibinfo{journal}{\bibinfo{title}{{Static and high frequency magnetic
  properties of FeGa thin films deposited on convex flexible substrates}}}.
\newblock {\emph{\JournalTitle{App. Phys. Lett.}}}
  \textbf{\bibinfo{volume}{106}}, \bibinfo{pages}{162405}
  (\bibinfo{year}{2015}).

\end{thebibliography}
\end{document}